\documentclass[intlimits,twoside,a4paper]{article}

\usepackage{amsmath,amssymb}
\usepackage{graphicx}

\usepackage[T2A]{fontenc}
\usepackage[cp1251]{inputenc}

\usepackage[eqsecnum]{cmpj2}

\usepackage{color}

\issue{2014}{17}{2}{23701}
\doinumber{10.5488/CMP.17.23701}

\title[Dispersion law and adiabatic potentials]%
{Similarities and differences in the construction of dispersion laws for charge carriers
in semiconductor crystals and adiabatic potentials in molecules}
\author{S.A.~Bercha, V.M.~Rizak}
\address{Uzhgorod National University, 54~Voloshyn St., 88000  Uzhgorod, Ukraine}

\date{Received January 17, 2014, in final form March 16, 2014}

\begin{document}

\maketitle

\begin{abstract}
Using the group theory and the method of invariants, it is shown how the vibronic potential can be
written in a matrix form and the corresponding adiabatic potentials can be found.
The molecule having $D_{3d}$ symmetry is considered herein as an example. The symmetries of normal vibrations active in Jahn-Teller's
effect were defined. $E$--$E$ vibronic interaction was considered to obtain vibronic potential energy
in a matrix form and thus the adiabatic potential.
Significant differences are shown in the construction of a secular matrix $D(\vec{k})$ for defining a
dispersion law for charge carriers in the crystals and the matrix of vibronic potential
energy, which depends on the normal coordinates of normal vibrations active in Jahn-Teller's effect.
Dispersion law of charge carriers in the vicinity of $\Gamma$ point of Brillouin
zone of the crystal with $D_{3d}^2$ symmetry was considered as an example.
\pacs  71.70.Ej, 71.20.-b, 61.50.Ah, 02.20.-a, 31.15.Hz
\end{abstract}

\section{Introduction}

The method of invariants was originally introduced in electronic theory of solids by Luttinger~\cite{ref1}
while considering a Bloch's electron in the magnetic field $\vec{H}$. Luttinger presented a secular matrix
$D(\vec{k},\vec{H})$, (where $\vec{k}$ is a small vector which starts at a high symmetry point of the
Brillouin zone with wave vector $\vec{k}_0$) as a sum of invariants. These invariants are products of
functions which depend on vector $\vec{k}$ components, intensity of magnetic field and operators of
the momentum, which are presented in a matrix form.

G.E.~Pikus had formalized the construction of secular matrices for defining the dispersion
laws for charge carriers using group theory methods~\cite{ref2}. He had introduced a concept of
basis matrices $A_{is}$, which can be built using the irreducible representations of $\vec{k}_0$ wave vector group.

Based on the group of $\vec{k}_0$ wave vector, in the vicinity of which one  considers the dispersion
law for charge carriers in~\cite{ref2}, formulas
were established to find irreducible representations $\tau_\mathrm{s}$. Basic matrices, basic functions which depend on $k_x$, $k_y$, $k_z$,
on the components of
the strain tensor and on the components of a magnetic field all transform according to above mentioned
representations $\tau_\mathrm{s}$.

Basic matrixes which are used in constructing a secular matrix $D(\vec{k})$, form well known sets of matrices.
For a double degenerated energy state in the $\vec{k}_0=0$ point, basic matrices are the Pauli's
matrices and the unit matrix of the second rank. Triple degenerated states  have got a set of basic matrices
which correspond to momentum operators $\hat{P}^j$ for $j=3/2$ written in a matrix form.

Thus, depending on the rank of a secular matrix $D(\vec{k})$, we have limited the number of the basic matrices.
Moreover, they can be the same for different irreducible representations
which describe crystals of different symmetry (space groups).

In this work we consider the possibility of applying the Pikus' method of invariants, formalized
in the group theory terminology, in order to determine the vibronic potential energy and adiabatic
potential in highly symmetric molecules.

We also analyze the similarities and differences in the construction of the vibronic potential energy
of molecules and the matrix of the dispersion law for a solid obtained in the center of the Brillouin zone.
For this reason, crystals with space group coinciding with the point group of a molecule are considered herein.

\section{Jahn-Teller's effect}

It is known that in highly symmetric molecules the Jahn-Teller's effect~\cite{ref3} is often observed.
This effect causes the reduction of symmetry of a molecule due to electron-vibronic interaction.
This interaction causes a split of a degenerated electronic term  and a change of the
configuration of a molecule. Energy of a molecule, as a function of the distance between the cores, should have a minimum
for a stable configuration. Obviously, it means that the expansion of the energy of a molecule by small
displacements of cores has no linear terms. Generally speaking, such terms appear when the adiabatic
approximation is broken due to the so-called vibronic interaction.

Any complicated movement of cores of a molecule can be represented as a series of harmonic oscillations.
Each of those is described by normal coordinates. The number of normal coordinates is equal to the number of degrees of freedom of a molecule.

Hamiltonian of an electronic subsystem now includes perturbational terms of vibronic potential energy.
This potential energy is tailored by normal displacements as follows~\cite{ref4}:
\begin{equation}
\label{eq1}
U = \sum_{\alpha,i}{V_{\alpha i}(q)Q_{\alpha i}}+\sum_{\alpha, i,\beta, k}{W_{\alpha i,\beta k}(q)Q_{\alpha i}Q_{\beta k}}+\cdots\, .
\end{equation}
As it was mentioned above, the linear part is the most important in Jahn-Teller's effect realization.

The first correction in perturbation theory  is defined by a matrix element:
\begin{equation}
\label{eq2}
V_{\rho\sigma} = \sum_{\alpha i}{Q_{\alpha i}\int{\Psi^*_{\rho}\hat{V}_{\alpha i}(q)\Psi_{\sigma}\,\rd q}},
\end{equation}
here, $\Psi_{\rho}$, $\Psi_{\sigma}$ are wave functions of a degenerate electron state and integration
is performed over the electronic configuration space $\{ q \}$.

From the invariance of the Hamiltonian which includes the linear $Q_{\alpha i}$ term it follows, that
the coefficient $V_{\rho\sigma}$ transforms by the elements of symmetry of a molecule in the same way as the
normal coordinates $Q_{\alpha i}$ do. In formulas (\ref{eq1}) and (\ref{eq2}), greek  indices
$\alpha$, $\beta$, $\ldots$ mean the number of irreducible representation, and $i$ and $k$~---
the number of base functions of this irreducible representation (taken in the form of normal coordinates).

It is known that  the secular equation, built on Hamiltonian $D(\vec{k})$ in a matrix form,
is used to find the dispersion law $E(\vec{k})$ for charge carriers in crystals.

The same procedure is used in case of a vibronic interaction in a molecule.
Adiabatic potential can be found after solving the secular equation that is built on a matrix of
the vibronic potential energy $D(Q_1,Q_2,\ldots)$. Here, $Q_1$, $Q_2$, $\ldots$~---  are normal
displacements of vibrations which are active in Jahn-Taller's effect.

In general, the adiabatic potential predicts that there can be several stable and metastable
configurations of a molecule.

There are some fundamental differences in constructing the matrix $D(\vec{k})$ and vibronic
interaction potential energy operator in a matrix form. The first one is the difference between
coefficients at the components of $\vec{k}$ wave vector and coefficients at the components of normal displacement.

The construction of $D(\vec{k})$ matrix which is used to find $E(\vec{k})$, is based on
$\vec{k}\cdot\hat{\vec{p}}$-approximation and on the method of  perturbation theory~\cite{ref5}.
It is obvious that coefficients of $D(\vec{k})$ matrix are integrals of  $\int{\Psi_i^*P_{\alpha}\Psi_j\,d\tau}$ type.
These expressions are of two kinds (let us denote them I and II, respectively) which corresponds
to the first and second perturbation corrections. In terms of type I, $P_{\alpha}$ is a component
of the operator of an impulse, $\Psi_i$ and $\Psi_j$ are functions that describe the
\emph{one degenerate electronic state}, for which the dispersion law $E(\vec{k})$ is investigated,
while in terms of type II, these functions belong to different states of a crystal.
This means that if we want to define whether the integral equals zero or not we should
investigate the antisymmetrized product of an irreducible representation that is built
only on functions of this degenerate electronic state. Antisymmetrization is connected
with the imaginary nature of the momentum operator $\hat{\vec{p}}$ (the perturbing part of $\vec{k}\cdot\hat{\vec{p}}$-approximation
includes the operator $\hat{\vec{p}}$)~\cite{ref5}.

Potential energy of vibronic interaction (\ref{eq1}) has two terms that are built on linear
combinations of components of normal displacements and their quadratic terms. Constructing
the matrix of potential energy of vibronic interaction [unlike the constructing of $D(\vec{k})$]
requires only the first correction of perturbation theory. It means that the matrix element of $D(\vec{k})$
matrix is built only on the eigenfunctions of the chosen degenerate state. Potential energy is an operator
of multiplication [$V_{\alpha i}$ and $W_{\alpha i}$ in equation (\ref{eq1})]. Matrix elements built
on the functions of a degenerate term of this operator will be evaluated by constructing the character of
a symmetrized product of irreducible representation that describes a degenerate electronic term.
Irreducible representation (denoted by $\tau_\mathrm{s}$) for transformation of the components of normal
displacements and their quadratic combinations is determined by equation~\cite{ref6}:
\begin{equation}
\label{eq3}
\frac{1}{2n}\sum_{g\in G}{\chi^\mathrm{s}(g)\left\{\left[\chi(g)\right]^2+\chi\left(g^2\right)\right\}}=1,
\end{equation}
where $\chi(g)$ is taken from the table of irreducible representations of a group of symmetry $G$ of a molecule.

The above statements contain the main differences in constructing the matrix of vibronic potential energy and the matrix $D(\vec{k})$.

Furthermore, unlike the matrix of vibronic potential energy, in order to construct a secular matrix which
consists of a sum of invariants (the product of basic matrixes and functions that depend on the components
of a wave vector $k_x$, $k_y$, $k_z$), one needs to consider not only equation (\ref{eq3}) but also the following formula~\cite{ref7, ref8}:
\begin{equation}
\label{eq4}
\frac{1}{2n}\sum_{g\in G}{\chi^\mathrm{s}(g)\left\{\left[\chi(g)\right]^2-\chi\left(g^2\right)\right\}}=1.
\end{equation}

Equation (\ref{eq3}) gives us $\tau_\mathrm{s}$ for even combinations of components of
a wave vector and equation (\ref{eq4}) provides $\tau_\mathrm{s}$ for odd combinations.
Basic matrices that form the $D(\vec{k})$ matrix are defined from the obtained~$\tau_\mathrm{s}$.

\section{Implementation of theory to ethane molecule}

The symmetry of the ethane molecule is described by $D_{3d}$ point group which has two-dimensional
irreducible representations. These representations correspond to double degenerated electronic
states (see table~\ref{tab1}). In a crystal belonging to a crystallographic class with the same point
symmetry $D_{3d}$, we will consider the group of the wave vector $\vec{k}_0=0$. In table~\ref{tab1}
both types of notations (i.e., molecular and for point $\Gamma$) for an irreducible representation are presented.
Also in table~\ref{tab2} we present the matrices of two-dimensional irreducible representations
$\Gamma_5$ and $\Gamma_6$ in the real form (unlike the complex one presented in the book by O.V.~Kovalev~\cite{ref10}).

\begin{table}[!b]
\caption{Characters of irreducible representations of a point group $D_{3d}$  and a group of
wave vector $\vec{k}=0$ for the space group $D_{3d}^2$ (denotation of elements of symmetry
is in correspondence with O.V. Kovalev, $h_{13}$ is the operation of inversion~\cite{ref10}).}
\label{tab1}
\vspace{2ex}
\begin{center}
\renewcommand{\arraystretch}{0}
\begin{tabular}{|c||c|c|c|c|c|c|}
\hline\hline
   &	$h_1$&	 $h_3$, $h_5$ &	 $h_8$, $h_{10}$, $h_{12}$&	 $h_{13}$&	 $h_{15}$, $h_{17}$&	 $h_{20}$, $h_{22}$, $h_{24}$ \strut\\
   \hline\hline
$A_g$, $\Gamma_1$ &	1&	 1&	 1&	 1&	 1&	 1 \strut\\ \hline
$A_u$, $\Gamma_2$ &	1&	 1&	 1&	-1&	-1& -1 \strut\\ \hline
$B_g$, $\Gamma_3$ &	1&	 1&	-1&	 1&	 1& -1 \strut\\ \hline
$B_u$, $\Gamma_4$ &	1&	 1&	-1&	-1&	-1&  1 \strut\\ \hline
$E_g$, $\Gamma_5$ &	2&	-1&	 0&  2&	-1&  0 \strut\\ \hline
$E_u$, $\Gamma_6$ &	2&	-1&	 0&	-2&  1&  0 \strut\\ \hline\hline
\end{tabular}
\renewcommand{\arraystretch}{1}
\end{center}
\end{table}

Thus, we will consider the so-called $E-E$ vibronic bonding, because the vibrational states will obviously
transform according to the same irreducible representations of $D_{3d}$ group.

To construct the vibronic potential energy matrix of the ethane molecule (C$_2$H$_6$) having a $D_{3d}$ point
symmetry we will define normal vibrations active in Jahn-Teller's effect. These normal vibrations should be
chosen from the following set: $3A_{1g}$, $1A_{1u}$, $2A_{2u}$, $3E_g$, $3E_u$~\cite{ref4}.

Calculations show that normal oscillations which are active in the
Jahn-Teller's effect have $A_{1g}$ and $E_g$ symmetry. Normal oscillation $A_{1g}$ should be excluded
whereas the configuration of the molecule does not change with such a normal displacement.

\begin{table}[!t]
\caption{Irreducible representations $\Gamma_5$ ($E_g$) and $\Gamma_6$ ($E_u$) written as real matrices.}
\label{tab2}
\vspace{2ex}
\begin{center}
\renewcommand{\arraystretch}{0}
\begin{tabular}{|c||c|c|c|c|c|c|}
\hline\hline
& $h_1$ & $h_3$ & $h_5$ & $h_8$ & $h_{10}$ & $h_{12}$ \strut\\ \hline
$\Gamma_5$ &
   $\begin{pmatrix}
     1 & 0 \\[1em]
     0 & 1 \\
   \end{pmatrix}$ &
   $\begin{pmatrix}
   -\frac12 & -\frac{\sqrt3}{2} \\[1em]
     \frac{\sqrt3}{2} & -\frac12 \\
   \end{pmatrix}$ &
   $\begin{pmatrix}
   -\frac12 & \frac{\sqrt3}{2} \\[1em]
   -\frac{\sqrt3}{2} & -\frac12 \\
   \end{pmatrix}$ &
    $\begin{pmatrix}
     1 & 0 \\[1em]
     0 &-1 \\
   \end{pmatrix}$ &
   $\begin{pmatrix}
   -\frac12 & \frac{\sqrt3}{2} \\[1em]
    \frac{\sqrt3}{2} & \frac12 \\
   \end{pmatrix}$ &
    $\begin{pmatrix}
   -\frac12 & \frac{\sqrt3}{2} \\[1em]
    \frac{\sqrt3}{2} & \frac12 \\
   \end{pmatrix}$  \strut \\ \hline \hline
  & $h_{13}$& $h_{15}$ & $h_{17}$ & $h_{20}$ & $h_{22}$ & $h_{24}$ \strut\\ \hline
$\Gamma_5$ &
   $\begin{pmatrix}
     1 & 0 \\[1em]
     0 & 1 \\
   \end{pmatrix}$ &
   $\begin{pmatrix}
   -\frac12 & -\frac{\sqrt3}{2} \\[1em]
     \frac{\sqrt3}{2} & -\frac12 \\
   \end{pmatrix}$ &
   $\begin{pmatrix}
   -\frac12 & \frac{\sqrt3}{2} \\[1em]
   -\frac{\sqrt3}{2} & -\frac12 \\
   \end{pmatrix}$ &
    $\begin{pmatrix}
     1 & 0 \\[1em]
     0 &-1 \\
   \end{pmatrix}$ &
   $\begin{pmatrix}
   -\frac12 & \frac{\sqrt3}{2} \\[1em]
    \frac{\sqrt3}{2} & \frac12 \\
   \end{pmatrix}$ &
    $\begin{pmatrix}
   -\frac12 & \frac{\sqrt3}{2} \\[1em]
    \frac{\sqrt3}{2} & \frac12 \\
   \end{pmatrix}$  \strut \\ \hline\hline
& $h_1$ & $h_3$ & $h_5$ & $h_8$ & $h_{10}$ & $h_{12}$ \strut\\ \hline
$\Gamma_6$ &
   $\begin{pmatrix}
     1 & 0 \\[1em]
     0 & 1 \\
   \end{pmatrix}$ &
   $\begin{pmatrix}
   -\frac12 & -\frac{\sqrt3}{2} \\[1em]
     \frac{\sqrt3}{2} & -\frac12 \\
   \end{pmatrix}$ &
   $\begin{pmatrix}
   -\frac12 & \frac{\sqrt3}{2} \\[1em]
   -\frac{\sqrt3}{2} & -\frac12 \\
   \end{pmatrix}$ &
    $\begin{pmatrix}
     1 & 0 \\[1em]
     0 &-1 \\
   \end{pmatrix}$ &
   $\begin{pmatrix}
   -\frac12 & \frac{\sqrt3}{2} \\[1em]
    \frac{\sqrt3}{2} & \frac12 \\
   \end{pmatrix}$ &
    $\begin{pmatrix}
   -\frac12 & \frac{\sqrt3}{2} \\[1em]
    \frac{\sqrt3}{2} & \frac12 \\
   \end{pmatrix}$  \strut \\ \hline \hline
  & $h_{13}$& $h_{15}$ & $h_{17}$ & $h_{20}$ & $h_{22}$ & $h_{24}$ \strut\\ \hline
$\Gamma_6$ &
   $\begin{pmatrix}
     -1 & 0 \\[1em]
     0 & -1 \\
   \end{pmatrix}$ &
   $\begin{pmatrix}
   \frac12 & \frac{\sqrt3}{2} \\[1em]
   -\frac{\sqrt3}{2} & \frac12 \\
   \end{pmatrix}$ &
   $\begin{pmatrix}
   \frac12 & -\frac{\sqrt3}{2} \\[1em]
   \frac{\sqrt3}{2} & \frac12 \\
   \end{pmatrix}$ &
    $\begin{pmatrix}
     -1 & 0 \\[1em]
     0 & 1 \\
   \end{pmatrix}$ &
   $\begin{pmatrix}
   \frac12 & -\frac{\sqrt3}{2} \\[1em]
   -\frac{\sqrt3}{2} & -\frac12 \\
   \end{pmatrix}$ &
    $\begin{pmatrix}
   \frac12 &-\frac{\sqrt3}{2} \\[1em]
    -\frac{\sqrt3}{2} & -\frac12 \\
   \end{pmatrix}$  \strut \\ \hline\hline
\end{tabular}
\renewcommand{\arraystretch}{1}
\end{center}
\end{table}

In case of  the so-called $E_g-E_g$ vibronic bonding in electron-vibrational interaction, there participate a double
degenerate electronic state with $E_g$  symmetry  and a normal oscillation of the molecule with $E_g$  symmetry .

Hence, the matrix of potential energy of vibronic interaction depends on two variables $Q_1$ and $Q_2$.
It can also include  squared combinations of $Q_1$   and $Q_2$. The method of projective operator was used
to find those squared combinations~\cite{ref11}. Calculations show that such combinations are functions $Q_1^2-Q_2^2$ and $2Q_1Q_2$.

The following matrices transform in accordance with representation $E_g$, i.e.,  basic matrixes $\sigma_x$
and $\sigma_z$, Pauli's matrices chosen from the set and  the identity matrix of the second rank $\sigma_1$.
Such a result is gained after applying matrix transformation rules under symmetry elements.

One can get a $D(Q_1,Q_2)$ matrix having  built the invariants from basic matrices and functions
\begin{equation} \label{eq5}
D(Q_1,Q_2)=\frac12\omega^2\left(Q_1^2+Q_2^2\right)\sigma_1+VQ_1\sigma_x+ WQ_1Q_2\sigma_x+VQ_2\sigma_z+W\left(Q_1^2-Q_2^2\right)\sigma_z \, ,
\end{equation}
here, $V$ and $W$  are coefficients of linear and quadratic parts of the operator of potential energy of
vibronic interaction, $\frac12\omega^2\left(Q_1^2+Q_2^2\right)$ is potential energy of normal oscillation
of a molecule which is described by $E_g$ representation, without getting vibronic interaction to account.

We should note that the same matrix of potential energy of vibronic interaction was obtained
by us in~\cite{ref12} for a molecule of methane (CH$_4$) whose symmetry is described  by the point group $C_{3v}$.

Despite the identical matrices of vibronic potential energy, normal displacements $Q_1$ and  $Q_2$ for
symmetric molecule C$_2$H$_6$ and non-symmetric molecule CH$_4$~\cite{ref12} differ significantly. The point is that
normal displacements $Q_1$ and $Q_2$ for C$_2$H$_6$ molecule are even functions, while in case of CH$_4$
molecule they have undefined parity. Even functions $Q_1$ and $Q_2$ are the base for irreducible
representation $E_g$ ($\Gamma_5$) of a point group $D_{3d}$. In case of CH$_4$ molecule
(a point group $C_{3v}$) $Q_1$ and $Q_2$ are the base for representation $E$ ($\Gamma_3$).

\section{Constructing the adiabatic potential}

It is known that adiabatic potential is defined by solving the secular equation constructed on the
matrix of vibronic potential energy. Adiabatic potential should reproduce the symmetry of a chosen molecule.
To fulfill this condition, one needs to replace the basic functions $Q_1$ and $Q_2$ of $E_g$ representation
with basic functions of an equivalent representation written in cartesian coordinates: $x^2-y^2$, $2xy$.
Due to the fact that $C_{3v}$  group does not include the operation of inversion, Cartesian coordinates $x$
and $y$ can additionally be the base for representation $E$ of this group (besides $x^2-y^2$ and $2xy$).

Thanks to the symmetrical matching of $Q_1$ and $Q_2$ with functions $x^2-y^2$, $2xy$, one can rewrite
the matrix of vibronic potential energy for $D_{3d}$ group in the form of a dependence on Cartesian coordinates.
Matrix elements of the above mentioned matrix are written as follows:
\begin{align} \label{eq6}
D_{11}&=\frac12 \omega^2\left(x^2+y^2\right)+2V_{xy}+W\left(x^4+y^4-6x^2 y^2\right),\nonumber \\
D_{22}&=\frac12 \omega^2 \left(x^2+y^2\right)-2V_{xy}-W\left(x^4+y^4-6x^2 y^2\right),	 \nonumber\\			
D_{12}&=D_{21}=V\left(x^2-y^2\right)+2W\left(x^2-y^2\right)2xy.
\end{align}
Such a denotation of the matrix makes possible the transformation to polar coordinates: $x=\rho\cos\varphi$, $y=\rho\sin\varphi$.

Having solved the secular equation obtained from $D(\rho,\varphi)$ matrix one gets the adiabatic potential:
\begin{equation}\label{eq7}
\varepsilon_{1,2}(\rho,\varphi)=\frac{\omega^2 \rho^4}{2} \pm \left[V\rho^4+2VW\rho^6 \sin6\varphi+ W^2 \rho^8\right]^{\frac12}.
\end{equation}
The presence of $\sin 6\varphi$ in the expression for adiabatic potential indicates the six minima in
its structure in contrast to the three minima in case of non-centrosymmetric molecule CH$_4$~\cite{ref12}.
The structure of adiabatic potential for the considered centrosymmetric molecule reflects its symmetry.

As a result of Jahn-Teller's effect, the lowering of symmetry can occur in two ways: the loss of centre of
symmetry or the loss of elements of symmetry (rotations $C_3$ and $C_3^2$). Namely, the lowering of symmetry
occurs from  $D_{3d}$ to $C_{3v}$ group or from $D_{3d}$ to $C_{2h}$ group.

Let us consider constructing the  secular matrix $D(\vec{k})$ in the vicinity of $\vec{k}_0=0$ for
a crystal having a $D_{3d}^2$ symmetry. We will choose an irreducible representation $E_g$ ($\Gamma_5$),
that describes a degenerate energy state. Such a choice is conditioned by the aim to analyze the similarities
in a matrix of potential energy of vibronic interaction of molecules and a secular matrix of the energy
spectrum of a crystal. According to Pikus' method of invariants, as it was mentioned, one can find $\tau_\mathrm{s}$-representation,
according to which the basic matrices as well as linear and square functions of the wave vector get transformed.

To find $\tau_\mathrm{s}$, the following equation is used~\cite{ref7}:
\begin{equation}\label{eq8}
n_\mathrm{s}=\sum_{g\in G_{\vec{k}_0}}|\chi(g)|^2\chi_\mathrm{s}(g).
\end{equation}
Trial characters $\chi_\mathrm{s}(g)$ are taken from the table of characters of a group of
wave vector  $\vec{k}_0=0$ (table~\ref{tab1}). Calculation shows that $n_\mathrm{s}\neq0$
only when $\tau_\mathrm{s}=\Gamma_1, \Gamma_3, \Gamma_5$.

Thus, we find that the basic matrices as well the $f(\vec{k})$  functions get converted
by representations $\Gamma_1$, $\Gamma_3$, $\Gamma_5$.

There is another significant difference in constructing a $D(Q_1,Q_2)$ matrix and a secular
matrix $D(\vec{k})$. From the selection rules [equations (\ref{eq3}) and (\ref{eq4})] one
gets different irreducible representations  $\tau_\mathrm{s}=\Gamma_1, \Gamma_3, \Gamma_5$
that describe the functions and basic matrices of invariants on which $D(\vec{k})$ matrices are built.
In case of constructing the matrix of potential energy of vibronic interaction,
only one normal displacement responsible for the vibronic interaction of electronic and vibronic states is chosen
out of all possible normal displacements that were gained from selection rules of irreducible representations.
In our case, it is the $E_g-E_g$ interaction.

We should note that irreducible representations $\tau_\mathrm{s}$, gained from equation (\ref{eq8})
should be redistributed to those that describe even and odd $f(\vec{k})$ functions. We use
equations (\ref{eq3}) and (\ref{eq4}) for this purpose.

Calculations show that a symmetrical squared character of representation $\Gamma_5$ [equation (\ref{eq8})]
includes representations  $\Gamma_1$ and $\Gamma_5$ while antisymmetrical one includes $\Gamma_3$ representation.
Thus, antisymmetrical function of a wave vector should be transformed by $\Gamma_3$ representation,
which is impossible (see table~\ref{tab1}).

Using the method of projective operator, one gets combinations of components of a wave vector that
correspond to representations $\Gamma_1$ and $\Gamma_5$.

It is clear that functions $k_x^2+k_y^2$, $k_z^2$ and an identity matrix are transformed by $\Gamma_1$
representation and functions $k_x k_z$ and  $k_y k_z$ are the base for representation $\Gamma_5$.

As it was shown before, in constructing the $D(Q_1,Q_2)$ matrix, the basic matrices that are transformed by
representation $E_g$  are $\sigma_x$ and $\sigma_z$. As representation $E$ matches $\Gamma_5$,
these matrices also correspond to representation $\Gamma_6$.

The calculated functions and matrices included in invariants are presented in table~\ref{tab3}.

\begin{table}[!t]
\caption{Matching the representations $\Gamma_1$ and $\Gamma_5$ with $f(\vec{k})$ functions and basic matrices
included in constructing the invariants.}
\label{tab3}
\vspace{2ex}
\begin{center}
\renewcommand{\arraystretch}{1}
\begin{tabular}{|c||c|c|c|}
\hline\hline
\raisebox{-1.7ex}[0pt][0pt]{representation} &\multicolumn{2}{c|}{$f(\vec{k})$\strut}
& \raisebox{-1.7ex}[0pt][0pt]{$A_{ls}$} \strut\\ \cline{2-3}
               &	$\gamma=1$       & $\gamma=-1$ &  \strut\\ \hline\hline
$\Gamma_1$     &$k_x^2+k_y^2$, $k_z$ &	--   &	 $\begin{pmatrix}
               1 & 0 \\
               0 & 1 \\
               \end{pmatrix}$ \strut\\ \hline
$\Gamma_5$     &$k_x k_z$, $k_y k_z$ &	--   &	 $\begin{pmatrix}
               1 & 0 \\
               0 &-1 \\
               \end{pmatrix}$,
               $\begin{pmatrix}
               0 & 1 \\
               1 & 0 \\
               \end{pmatrix}$ \strut\\ \hline\hline
\end{tabular}
\renewcommand{\arraystretch}{1}
\end{center}
\end{table}

Based on data from table~\ref{tab3}, we construct a secular matrix $D(\vec{k})$:
\begin{equation}\label{eq9}
D(\vec{k})=\begin{pmatrix}
             a\left(k_x^2+k_y^2\right)+bk_z^2+ ck_x k_z & ck_y k_z \\
             ck_y k_z & a\left(k_x^2+k_y^2\right)+bk_z^2 - ck_x k_z \\
           \end{pmatrix}.
\end{equation}

By solving a corresponding secular equation we obtain an expression for the dispersion law of
charge carriers in point $\vec{k}_0=0$ for the state described by $\Gamma_5$ representation:
\begin{equation}\label{eq10}
E(\vec{k})=a\left(k_x^2+k_y^2\right)+bk_z^2\pm\sqrt{c^2 k_z^2\left(k_x^2+k_y^2\right)}\,.
\end{equation}

From equations (\ref{eq7}) and (\ref{eq10}), we conclude that solutions of corresponding secular
equations reproduce the point symmetry of the crystal and the molecule.

\section{Conclusions}

Thus, the secular matrix $D(k)$ as well as the matrix of vibronic potential energy are built from the sum of invariants.
In both cases, each of these invariants  is a product of the basis matrix (Pauli's matrices in our case)
and the basis function which depends on corresponding variables.

In the case of a secular matrix, basis functions and basis matrices transform according to irreducible
representations, which form symmetrized and antisymmetrized squares of the irreducible representation
connected with an active normal vibration or with a corresponding degenerated electronic term for which the
secular matrix is written down. Polynomials from which the basis functions are built are powers of wave
vector's components. This small wave vector originates from point in the Brillouin zone in the vicinity
of which one construct the dispersion law $E(k)$.

In the case of vibronic potential energy construction, the basis matrices and functions are built solely
for irreducible representations which form a symmetrized square of the irreducible representation,
describing the vibration which is active in Jahn-Teller's effect. Corresponding basis functions are also
built on components of this vibration.

In conclusion, we should note that the construction of vibronic potential energy and the adiabatic
potential can be achieved without using the  method of  invariants, solely by using the Clebsch-Gordan
coefficients~\cite{ref13}. A correct solution of the adiabatic potential construction problem by means of
group theory method and the method of invariants allows one to successfully apply adiabatic potentials for a qualitative
explanation of a wide variety of phenomena connected with a vibronic interaction in molecules and crystals.
Moreover, the method of adiabatic potential construction can be adapted to the investigation of peculiarities of phase
transitions in crystals with Jahn-Teller centers (for example, the CuInP$_2$S$_6$ crystal~\cite{ref9}).
The mentioned problem will be investigated in our next work.

\section*{Acknowledgements}

Author (B.S.A.) wishes to thank Dr. Glukhov K.E. for helpful discussions and critical reading of the manuscript.


\newpage

\ukrainianpart

\title{Спільність і відмінність у побудові  законів дисперсії носіїв заряду  в напівпровідникових кристалах і адіабатичних потенціалів у молекулах}
\author{С.А. Берча, В.М. Різак}
\address{ДВНЗ ``Ужгородський національний університет'', вул. Волошина, 54, 88000 Ужгород, Україна}

\makeukrtitle

\begin{abstract}
\tolerance=3000%
У роботі показано як, використовуючи  теоретико-груповий метод і метод інваріантів, можна одержати вібронний потенціал, записаний у матричному вигляді, та відповідні адіабатичні потенціали. В якості прикладу розглядається молекула з симетрією $D_{3d}$. Визначено симетрію нормальних коливань, активних в ян-теллерівському ефекті. Розглянуто $E-E$ вібронний зв'язок для одержання вібронної потенціальної енергії у матричному вигляді та адіабатичний потенціал.
Вказано на істотні відмінності у побудові секулярної матриці $D(\vec{k})$ для знаходження закону дисперсії електронного спектру в кристалах і матриці вібронної потенціальної енергії, залежної від нормальних координат активного в ян-теллерівському ефекті нормального коливання.
В якості прикладу розглядується закон дисперсії носіїв струму в околі точки $\Gamma$ зони Бріллюена кристалу з симетрією $D_{3d}^2$.
\keywords ефект Яна-Теллера, метод інваріантів, закон дисперсії, адіабатичний потенціал
\end{abstract}

\end{document}